\documentclass[11pt]{article}
\usepackage{mathptmx}
\usepackage{amssymb,amsmath,amsthm}

\DeclareMathOperator{\C}{\mathit{C}\,}

\newcommand{\RIGHT}{\rm RIGHT}
\newcommand{\LEFT}{\rm LEFT}
\newcommand{\zo}{\{0,1\}}
\newcommand{\poly}{{\rm poly}}
\newtheorem{definition}{Definition}
\newtheorem{theorem}{Theorem}

\newtheorem{lemma}[theorem]{Lemma}

\if01
\usepackage{fancyheadings}
\fi

\begin{document}

\title{Short lists with short programs in short time \\
 - a short proof}
\author{ {Marius Zimand\/}
\thanks{  Department of Computer and Information Sciences, Towson University,
Baltimore, MD.; email: mzimand@towson.edu; http://triton.towson.edu/\~{ }mzimand.
This work has been supported
by NSF grant CCF 1016158. }}

%\date{\normalsize \today}
\date{}

\maketitle  

\begin{abstract} Bauwens, Mahklin, Vereshchagin and Zimand~\cite{bmvz:t:shortlist} and Teutsch~\cite{teu:t:shortlist} have shown that given a string $x$ it is possible to construct in polynomial time a list containing  a short description of it. We simplify their technique and present a shorter proof of this result.
\end{abstract}

\section{Introduction}

Given that the Kolmogorov complexity is not computable, it is natural to ask if given a string $x$ it is posible to construct a short list containing a minimal (+ small overhead) description of $x$. Bauwens, Mahklin, Vereshchagin and Zimand~\cite{bmvz:t:shortlist} and Teutsch~\cite{teu:t:shortlist} show that, surprisingly,   the answer is YES.  Even more, in fact the short  list can be computed in polynomial time. More precisely,~\cite{bmvz:t:shortlist}  showed that one can effectively compute lists of quadratic size guaranteed to contain a description of $x$  whose size is  additively $O(1)$ from a minimal one (it is also shown  that it is impossible to have such lists shorter than quadratic),  and  that one can compute in polynomial-time lists guaranteed to contain a description that is additively $O(\log n)$ from minimal. Finally,~\cite{teu:t:shortlist} improved the latter result  by reducing  $O(\log n)$ to  $O(1)$.
\begin{theorem}[\cite{teu:t:shortlist}]
\label{t:poly}
For every standard machine $U$ there is  a constant $c$ and a polynomial-time algorithm $f$ such that for every $x$,  $f(x)$ outputs a list of programs
that contains a $c$-short program for $x$.\footnote{It can be shown that the list size is $n^{6 + \delta}$ for any arbitrarily small positive  constant $\delta$.} 
\end{theorem}
Let us explain the formal terms.
Given a  Turing machine $U$, a $c$-short program for $x$ is a string $p$ such that $U(p)=x$ and the length of $p$ is bounded by $c$+ (length of a shortest program for $x$). A  machine  $U$  is \emph{optimal} if $\C_U(x \mid y) \le C_V(x \mid y) + O(1)$ for all machines $V$  (where $C$ is the Kolmogorov complexity and the
constant $O(1)$ may depend on $V$).   
An optimal machine $U$ is \emph{standard}
if for every machine $V$ there is an efficient translator from any machine $V$ to $U$, i.e., a
polynomial-time  computable function $t$ such that for all $p,y$, $U(t(p),y) = V(p,y)$ and $|t(p)| = |p| +O(1)$.

Both~\cite{bmvz:t:shortlist} and ~\cite{teu:t:shortlist} prove their results regarding polynomial-time computable lists as corollaries of somewhat more general theorems.  We present in this note a direct proof of Theorem~\ref{t:poly}, which is simpler and shorter than the one in~\cite{teu:t:shortlist}. We emphasize that there is no technical innovation in the proof that we present below. We use the same general approach and the same ingredients as in~\cite{bmvz:t:shortlist} and ~\cite{teu:t:shortlist}, but,  because we go straight  to the target, we can take some shortcuts that render the proof simpler.\footnote{The proof given here also produces a smaller value for the constant in the theorem.}

\textbf{Proof overview.} Essentially we want to compress in polynomial time to (close to) minimal length, such that decompression is computable (not necessarily in poynomial time). This is of course impossible in absolute terms, but here we compress in a weaker sense, because we obtain not a single compressed string, but a list guaranteed to contain the (close to) optimally  compressed string. It is natural to think to use seeded extractors, because an extractor's  output is close to being optimally compressed in the Shannon entropy sense. The problem is that  we need an extractor with logarithmic seed (because we want a list of polynomial size) and no entropy loss (because we want to decompress). Unfortunately,  such extractors have not yet been shown to exist. The key observation from~\cite{bmvz:t:shortlist}, also used in~\cite{teu:t:shortlist}, is that in fact a disperser is good enough, and then one can use the disperser from~\cite{ats-uma-zuc:j:expanders}, which has the needed parameters.  Now, why are dispersers sufficient?  The answer, inspired by~\cite{mus-rom-she:j:muchnik},  stems from the idea from~\cite{bmvz:t:shortlist} to use for this kind of compression graphs that allow on-line matching.  These are unbalanced bipartite graphs, which, in their simplest form, have $\LEFT = \zo^n, \RIGHT=\zo^{k + \mbox{small overhead}}$, and left degree = $\poly(n)$, and  which permit on-line matching up to size $K=2^k$. This means that any set  $A$ of $K$  left nodes,  each one requesting   to be matched to some adjacent right node, can be satisfied in the on-line manner(i.e., the requests arrive one by one and  each request is satisfied before seeing the next one; in our proof we will allow a small number of requests to be discarded, but this should also happen before the next request arrives).  The correspondence to our problem is roughly that strings in LEFT are the strings that we want to compress, and the strings in RIGHT  are their compressed forms. We need on-line matching because we are going to enumerate left strings as they are produced by the universal machine and each time a string is enumerated we want to find it a match, i.e., to compress it.  In order for a graph to allow matching, it needs to have good expansion properties. It turns out that it is enough if left subsets of a given  size $K/O(1)$ expand to size $K$, and a disperser has this property. When we decompress, given the right node (the compressed string), we run the matching algorithm and see which left node has been matched to it. For this the decompressor  needs  to have $n$ to be able to construct the graph, and this produces the $O(\log n)$ overhead.  Thus  this approach  is good enough to obtain the result with $O(\log n)$-short programs from~\cite{bmvz:t:shortlist}. To reduce $O(\log n)$ to $O(1)$, we need the new ideas from~\cite{teu:t:shortlist}. The point is that this time we want LEFT to have strings not of a single length $n$, but of all lengths $n \geq k$ (because we can no longer afford  to give $n$ to the decompressor).  In fact, it is not hard to see, that  it is enough to  restrict to lengths $k \leq n \leq 2^{k}$. This time we need expansion for all sets of size $\leq K$ (not just equal to a fixed $K/O(1)$, because we need each subset (of the  match-requesting set $A$)  of strings of  a given length to expand. For this, the unbalanced lossless expander from~\cite{guv:j:extractor} is good, except for one problem: The size of  RIGHT in this expander  is $\poly(K)$ and not the desired  $K + O(1)$. This problem is fixed by compressing using again the disperser from~\cite{ats-uma-zuc:j:expanders} to a set of size $K \cdot \poly(k)$, and, finally, using a simple trick,  to size $K + O(1)$, which implies the  $O(1)$ overhead we aim for.

\section{The proof}

We use bipartite graphs $G = (L, R, E \subseteq L \times R)$. We denote  $\LEFT(G) = L$, $\RIGHT(G) = R$. For integers $ n,m, k,d$ we denote  $N=2^k, M=2^m, K=2^k, D= 2^d$. We denote $[n] = \{1,2, \ldots, n\}$.
A bipartite graph $G$  is explicit if there exists a polynomial-time algorithm that given $x \in \LEFT(G)$ and $i$, outputs the $i$-th neighbor of $x$.

\begin{definition}
A bipartite graph $G$ is a $(K,K')$-expander if every subset of left nodes having size $K$, has at least $K'$ right neighbors.
\end{definition}
\begin{theorem}[Guruswami, Umans, Vadhan~\cite{guv:j:extractor}] For every constant $\alpha$, every $n$, every $k \leq n$, and $\epsilon > 0$, there exists am explicit $(K', (1-\epsilon)DK')$ expander for every $K' \leq K$, in which every left node has degree  $D = O((nk/\epsilon)^{1+1/\alpha})$, $L = [N], R= [M]$, $M \leq D^2 \cdot K^{1+\alpha}$.
\end{theorem}
\begin{definition}
A bipartite graph $G = (L,R,E)$ is a \emph{$(K, \delta)$-disperser}, if every subset  
$B \subseteq L$ with $|B| \geq K$  has at least $(1-\delta)|R|$ distinct neighbors.
\end{definition}
\begin{theorem} [Ta-Shma, Umans, Zuckerman~\cite{ats-uma-zuc:j:expanders}] 
\label{t:tuz}
For every $K,n$ and constant $\delta$,  
there exists explicit $(K, \delta)$-dispersers 
$G = (L=\zo^n, R= \zo^m, E \subseteq L \times R)$ in which every node in $L$ has degree 
$D= n 2^{O((\log \log n)^2)}$ and $|R| =\frac{ \alpha K D}{n^3}$, for some constant $\alpha$.\footnote{\cite{ats-uma-zuc:j:expanders} only indicates that $D = \poly(n)$. The value  $D= n 2^{O((\log \log n)^2)}$ is obtained by reworking the proof in Lemma 6.4~\cite{ats-uma-zuc:j:expanders} using the extractor with constant loss from Theorem 4.21 in~\cite{guv:j:extractor}.}
\end{theorem}
The key combinatorial object that we use is provided in the following lemma.
\begin{lemma}
\label{l:cexpander}
For every constant $c$ and every sufficiently large $k$, there exists an explicit  bipartite graph $H_k$ with the following properties:
\begin{enumerate}
\item  $\LEFT(H_k)  = \zo^k \cup \zo^{k+1} \cup \ldots \cup \zo^{2^k}$,   $ \RIGHT(H_k) = \zo^{k+1}$,
%\item   $ \RIGHT(H_k) = \zo^{k+1}$,
\item  Each left  node $x$ has degree $\poly(|x|)$,
\item $H_k$ is a $(K/c^2, K)$-expander.
\end{enumerate}
\end{lemma}
We defer the proof of this lemma for later.  

We show how the lemma implies Theorem~\ref{t:poly}. We start with the following lemma about on-line matching (recall that this means that one receives a sequence of requests to match  left nodes  with one of their adjacent right nodes and
each request must be satisfied, or discarded, before seeing the next one).  
\begin{lemma}
\label{l:matchreject}
 If $K$  on-line matching requests are made in a  $(K/c^2,K)$-expander all but less than $K/c^2$  can be  satisfied. 
\end{lemma} 
\begin{proof}
Suppose there are $K$ requests for matching left nodes and we attempt to satisfy them in the obvious greedy manner.  Suppose that $K/c^2$ requests cannot be satisfied (because all their neighbors have been used to match previous requests). The $K/c^2$ left nodes that are not satisfied have $K$ right neighbors and all of them have satisfied
matching requests. This would imply that all the $K$ requests have been satisfied, contradiction.
\end{proof}

\textbf{Proof of Theorem~\ref{t:poly}}.
\medskip

We define the following machine $V$ (``the decompressor").
\medskip

\fbox{
\vbox{

(1) On inputs of the form $00p$, $V$ outputs $p$.

(2) On inputs of the form $01p$, $V$ simulates $U(p)$ and if $U(p)=x$ and $|x| > 2^{|p|}$, outputs $x$.

(3)  On inputs of the form $1p$, $V$ works as follows: 

 $V$ calculates its value on all inputs of the form $1p'$ with $|p'| = |p|$ as follows.  Let $k = |p|-1$. Enumerate the elements of the set $\{x \mid \exists q \mbox{ of length } k, U(q) = x\}$. When an element $x$ is enumerated and $|x|$ is between $k$ and $2^k$,
pass $x$ to the online matching algorithm for $H_k$. If $x$ is matched to $p'$,  then $V(p')$ outputs $x$. If $x$ is rejected because all its right neighbors in $H_k$ have already been used to match other elements during the computation of  $V(1p')$ for strings $p'$ of length $k-1$, continue the enumeration.
}
}
\medskip

Observe that during computations of the form (3), at most $K$ matching requests are made and therefore, by the property of $H_k$,  there are fewer than $K/c^2$ rejections. It follows that if $v$ is a rejected node then $C_U(v) \leq k - 2 \log c + \log c + 2 \log \log c +O(1) < k$, for $c$ a large enough constant. Indeed a rejected string can be described by its index in the set of rejected strings written on exactly $k- 2\log c$ bits, and $c$ (which is needed in order to reconstruct $k$ and next enumerate the set of rejected strings). The additional $2 \log \log c$ term is required for concatenating the index and $c$.   It follows that if $x$ is a string such that $C_U(x) = k$ and $k \in\{\log |x|, \ldots, |x|\}$, then there exists $p$ of length $k+1$ such that $V(1p)=x$. Moreover, $p$ is one of the right  neighbors of $x$ in $H_k$.

Now,  for each $x$, let $list(x)$ be the   list containing the following strings: $00x$, all strings of length $< \log |x|$ prefixed with $01$, and all the neighbors of $x$ in $H_k$ prefixed with a $1$, for $k=|x|, |x|-1, \ldots, \log(|x|)$.  Note that for every $x$, $list(x)$ can be computed in polynomial time, and there exists $v \in list(x)$, $|v| \leq C_U(x) + O(1)$ such that $C_V(v) = x$. Finally, using the "translator" $t$ from $V$ programs to $U$ programs, take $f(x) = \{t(v) \mid v \in list(x)\}$. Since $t$ is computable in polynomial time, $U(t(v)) = V(v)$ and $|t(v)| = |v| + O(1)$, we are done.~\qed
 \medskip

It remains to prove Lemma~\ref{l:cexpander}. We use two types of graphs given in the following two lemmas.
\begin{lemma}
\label{l:guv}
For every $n$, and $k \leq n$, there exists a bipartite graph $GUV_{n,k}$ with each left node having degree $D = \lambda (nk)^2$ (for some fixed constant $\lambda$), $\LEFT(GUV_{n,k}) = \zo^n$, $\RIGHT(GUV_{n,k}) = [M]$ with $M \leq  D^2K^2$ , which is a $(K', (1/2)DK')$-expander for every $K' \leq K$.
\end{lemma}
\begin{proof}
This is the Guruswami, Umans, Vadhan expander with parameters $\alpha = 1, \epsilon = 1/2$.
\end{proof}
\begin{lemma}
\label{l:fivek}
For every $k$, there exists a bipartite graph $F_k$ with each left  node having degree $D= O(k^3)$, $\LEFT(F_k) = \zo^{8k}$, $\RIGHT(F_k) = \zo^{k+1}$, which is a $(K,K)$-expander.
\end{lemma}
\begin{proof}
Consider the Ta-Shma, Umans, Zuckerman $(K, 1/2)$-disperser $G$, with $\LEFT(G) = \zo^{8k}$, $\RIGHT(G) = \zo^m$ ,  left degree $D = O(k 2^{O((\log \log k)^2)})$ and $|\RIGHT(G)| = \frac{\alpha KD}{(8k)^3}$.

To increase the size of the right set to be at least $2K$, we make  $\RIGHT$  consist of  $2 \lceil \frac{(8k)^3}{\alpha D} \rceil$ copies of  $\RIGHT(G)$ connected to $\LEFT(G)$ in the same way as the original nodes.  Thus each right node is labelled by a string of length $\geq k+1$ and the left degree is $O(k^3)$.

By merging the nodes whose labels have the same prefix of length $k+1$, we obtain the graph $F_k$, which as desired has   $\RIGHT(F_k) = \zo^{k+1}$ and  is a $(K, 1/2)$-disperser (because the merge operation can only improve the dispersion property).

Thus, every left subset of size $K$ has at least $(1/2) \cdot 2K$ right neighbors, i.e., $F_k$ is a $(K,K)$-expander.
\end{proof}
We are now prepared to prove Lemma~\ref{l:cexpander}.
\medskip

\textbf{Proof of  Lemma~\ref{l:cexpander}}
\medskip

Let us fix $c$ and a sufficiently large $k$.

 We first construct the graph $G_k$ as the union $GUV_{k,k} \cup  GUV_{k+1,k} \cup \ldots \cup GUV_{2^k,k}$.

Note that $\LEFT(G_k)$ consists of all strings having length between $k$ and $2^k$. For $\RIGHT(G_k)$, we shift the numerical labels of the right nodes in each set  in the obvious way before taking the union, so that the sets that we union are pairwise disjoint.  We have 
\[
|\RIGHT(G_k)| \leq  \sum_{n=k}^{2^k} \lambda^2 (nk)^4 K^2 = \lambda^2 k^4 K^2  \sum_{n=k}^{2^k} n^4 \leq \lambda^2 k^4 \cdot K^7 < K^8,
\]
for $k$ sufficiently large. By padding each right node in $G_k$ with $100 \ldots 0$, we label each right node by a string of length $8k$.

Note that, provided $k$ is sufficiently large, $G_k$ is a $(K/c^2, K)$-expander. Indeed take $B \subseteq \LEFT(G_k)$, $|B| = K/c^2$. $B$ has strings of different lengths.  If we partition $B$ into subsets of strings  corresponding to the different lengths, each subset with strings of length say $n$ expands according to $GUV_{n,k}$ by a factor of $(1/2) \lambda (nk)^2 \geq c^2$ (if $k$ is large enough). Since different subsets of the partition map into disjoint right subsets, the above assertion follows.

The degree of every left node $x$ in $G_k$ is bounded by $\poly(|x|)$ because the edges originating in $x$ are those from the graph $GUV_{|x|,k}$.  So $G_k$  is almost what we need except that the right nodes have length  $8k$ instead of $k+1$. We fix this issue by compressing strings of length $8k$ to length $k+1$ using the graph $F_k$ from Lemma~\ref{l:fivek}. 

More precisely, we build the graph $H_k$ by taking the product of the above graph $G_k$ with the graph $F_k$. Thus $\LEFT(H_k)  = \LEFT(G_k)$, $\RIGHT(H_k) = \RIGHT(F_k)$ and $(x,y)$ is an edge in $H_k$ if there exists $z \in \RIGHT(G_k) \subseteq  \LEFT(F_k)$ such that  $(x,z)$ is an edge in $G_k$ and $(z,y)$ is an edge in $F_k$. As desired, $\LEFT(H_k)$ consists of all strings $x$ having length between $k$ and $2^k$, $\RIGHT(H_k) = \zo^{k+1}$, the degree of every left node $x$ is bounded by $\poly(|x|)\poly(k) = \poly(|x|)$ and
$H_k$ is a $(K/c^2,  K)$-expander, because each left subset of size $K/c^2$ expands to size at least  $K$ in  $G_k$ and then it keeps its size at least  $K$ when passing through $F_k$.~\qed
\medskip

\textbf{Note.} The above construction yields in Theorem~\ref{t:poly} a list of size $O(n^8)$. If in  Lemma~\ref{l:guv} we take a small $\alpha$  (instead of $\alpha = 1$), we obtain list size $n^{6 + \delta}$, for arbitrarily small positive constant $\delta$.

\section{Acknowledgements} We are grateful to Alexander Shen for his comments and for signalling an error  in an earlier version.  We thank Jason Teutsch for useful conversations that lead to a more precise estimation of the list size in Theorem~\ref{t:poly}.
\if01

\section{Online matching}

\begin{theorem}
\label{t:online}
For every positive integer $k$, there is a bipartite graph $G$ with $LEFT(G) = \zo^{\geq k}$, $RIGHT(G) = \zo^{k+1}$,  $deg(x) = \poly (|x|)$ for every left node $x$, which has on-line matching up to $K$.  (Here $\poly|(x|)$ is a fixed polynomial that does not depend on $k$).
\end{theorem}
\begin{proof}
We prove the statement for every $k$ greater than some constant $k_0$ which will be defined later (it can be arranged that $k_0 = 4$ at the cost of increasing the polynomial that bounds the left degree). For $k < k_0$ one can just take the complete bipartite graph whose right hand side consists of all strings having length $k$.

So now we can consider that  $k$ is large enough for the following arguments to be valid. As we did in the previous proof, we take $G_k$ as the union $GUV_{k,k} \cup  GUV_{k+1,k} \cup \ldots \cup GUV_{2^k,k}$ but we prefix all the right nodes with $1$ and, as we did in the previous proof,  we append $100 \ldots 0$ so that all right nodes have length $8k+1$. By the same argument as in the previous proof, if $k$ is large enough, the graph  $G_k$ is a $(K'/8, K')$ expander for every $K' \leq  K$,  and $LEFT(G_k) = \zo^{[k \ldots 2^k]}$,  $RIGHT(G_k) = 1 \zo^{8k}$ and every left node $x$ has degree bounded by a fixed polynomial in $|x|$.
 
To handle nodes that have length greater than $2^k$ we consider the complete bipartite graph $A_k$ with $LEFT(A_k) = \zo^{>2^k}$ and $RIGHT(A_k) = 0 \zo^{8k}$ and consider the graph $G'_{k}$ obtained by taking the union of  $G_k$ and $A_k$. Note that $G'_k$ is a $(K'/8, K')$ expander for every $K' \leq  K$,  $LEFT(G'_k) = \zo^{\geq k}$,  $RIGHT(G_k) =  \zo^{8k+1}$ and every left node $x$ has degree  bounded by a fixed polynomial in $|x|$.

We next use the  Ta-Shma-Umans-Zuckerman disperser with $\epsilon = 1/3$ with the set of left nodes consisting of all $(8k+1)$ long strings. This time viewing the set of right nodes as an initial segment of the natural numbers, we partition the right nodes into $K + \lfloor 2K/3 \rfloor$ equal intervals and then we merge the right nodes in each interval (this is the analog of the merging operation that we did in the previous proof when  we merged the  nodes that have the same $k+1$ prefix).   We obtain a graph $F_k$  such that  $LEFT(F_k) = \zo^{8k+1}, 
 |RIGHT(F_k)| = K + \lfloor 2K/3 \rfloor$ and that is a $(K, 1/3)$ disperser. Thus every left subset of size $K$ has at least $(1 - 1/3)(K + \lfloor 2K/3 \rfloor) \geq K$ right neighbors  (for $K \geq 6$),  i.e., $F_k$ is a $(K,K)$ expander.

We take $H_k$ to be the product of $G'_k$ and $F_k$.  There exists a constant $k'_0$ such that if $k \geq k'_0$, then $H_k$ is a $(K/8, K)$ expander, with $LEFT(H_k) = \zo^{\geq k}$ and $|RIGHT(H_k)| = K + \lfloor 2K/3 \rfloor \leq 5K/3$. Also the degree of every left $x$ is bounded by a fixed  polynomial in $|x|$.

Next we take $G$ to be the union of the graphs $H_k \cup H_{k-3} \cup H_{k-6} \cup  \ldots \cup   H_{k'}$ where $k'$ is the smallest number of the form $k-3t$ which is at least $k'_0$. When we make the union, we keep the right sets disjoint but $LEFT(G) = \zo^{\geq k}$. Clearly, the degree of every left $x$ is bounded by a  fixed polynomial in $|x|$.   Suppose that we have to satisfy $K$ online matching requests. We attempt to satisfy them in the greedy manner, first using the edges in $H_k$, then the edges in $H_{k-3}$, and so on. By Lemma~\ref{l:matchreject},  less than $K/8$ are rejected at the first level $H_{k}$, less than $K/8^2$  remain rejected after the second level $H_{k-3}$, and so on.  It follows that less than  $2^{k'}/8$ requests  are still not satisfied when we reach the last level $H_{k'}$. Note that $k'$ is one of the numbers $k'_0, k'_0 +1$, or $k'_0 +2$.  By adding to the right set of $G$,  $2^{k'}/8$ new nodes which we connect to all left nodes (thus all left degrees increase by only a constant), we can satisfy the remaining requests.

The number of nodes in $G$ is bounded by 
\[
(5/3)K + (5/3)(K/8) + (5/3) (K/8^2) +   \ldots + (5/3)(K'/8) + 2^{k'}/8 \leq (5/3)(8K/7)  + 2^{k'_0+2}/8  \leq 2K,
\] if $k \geq k'_0 + 3$, a value which we take to be the constant $k_0$ mentioned at the beginning of the proof. Therefore we can label the right nodes in $G$ by strings of length $k+1$.
 \end{proof}
\fi
\bibliography{c:/book-text/theory}

\bibliographystyle{alpha}

\end{document}